\documentstyle[12pt]{article}

\begin{document}

\hfill{UTTG-04-04}

\vspace{36pt}

\begin{center}
{\large {\bf {Must Cosmological Perturbations Remain Non-Adiabatic After Multi-Field Inflation? }}}

\vspace{36pt}
Steven Weinberg\footnote{Electronic address:
weinberg@physics.utexas.edu}\\
{\em Theory Group, Department of Physics, University of
Texas\\
Austin, TX, 78712}

\vspace{30pt}

\noindent
{\bf Abstract}
\end{center}

Even if non-adiabatic perturbations are generated in multi-field inflation, the perturbations will become adiabatic if the universe after inflation enters an era of local thermal equilibrium, with no non-zero conserved quantities, and will remain adiabatic as long as the wavelength is outside the horizon, even when local thermal equilibrium no longer applies.  Small initial non-adiabatic perturbations associated with imperfect local thermal equilibrium remain small when baryons are created from out-of-equilibrium decay of massive particles, or when  dark matter particles go out of local thermal equilibrium.

\noindent

 \vfill

\pagebreak
\setcounter{page}{1}

\begin{center}
{\bf I. INTRODUCTION}
\end{center}

\vspace{12pt}

A recent paper [1] has shown, contrary to what is sometimes thought [2], that if the cosmological perturbations produced during inflation are in the adiabatic mode after they leave the horizon, then they stay in that mode through reheating and until horizon re-entry.
This paper considers a complementary question: if the cosmological perturbations produced during 
multi-field inflation are {\em not} in the adiabatic mode after they leave the horizon, then must they remain non-adiabatic thereafter?

As in [1], our approach to this problem is based on the theorem that in all cosmological models, among the various solutions of the differential equations that govern  perturbations in Newtonian gauge, there are always two independent solutions that are adiabatic throughout the whole period when the perturbations are outside the horizon.[3]  By ``adiabatic'' is usually meant that
the perturbation $\delta {\cal S}$ in any four-scalar ${\cal S}$ (including temperature, proper number densities, scalar fields, etc.) is proportional to $\dot{\bar{{\cal S}}}$, where bars denote unperturbed quantities, and dots denote ordinary time derivatives.  In particular, for the energy densities and   pressures of all the individual constituents (labeled $\alpha$) of the universe, we have[4]  
\begin{equation}
\frac{\delta\rho_\alpha  }{\dot{\bar{\rho}}_\alpha}=\frac{\delta p_\alpha  }{\dot{\bar{p}}_\alpha} \;\;{\rm is\;independent\;of}\;\alpha\;,
\end{equation} 
from which comes the term ``adiabatic.''
The theorem of ref. [3] provides the further information that for all four-scalars ${\cal S}$,  the adiabatic solution has
\begin{equation} 
\frac{\delta S}{\dot{\bar{\cal S}}}=-{\cal I}\;,
\end{equation} 
and in particular 
\begin{equation}
\frac{\delta\rho_\alpha  }{\dot{\bar{\rho}}_\alpha}=\frac{\delta p_\alpha  }{\dot{\bar{p}}_\alpha} =-{\cal I}\;,
\end{equation}
where 
\begin{equation}
{\cal I}(t)\equiv \frac{\zeta}{a(t)}\int_{\cal T}^t a(t)dt\;.
\end{equation} 
Here $a(t)$ is the Robertson--Walker scale factor, and $\zeta$ and ${\cal T}$
are constants, the same for all four-scalars.  (The freedom to choose ${\cal T}$ as well as $\zeta$ means that these are two independent solutions.)  Also, according to this theorem, the three-scalar gravitational
potentials (defined by the Newtonian gauge line element $ds^2=-(1+2\Phi)dt^2+a^2(1-2\Psi)d{\bf x}^2$) are [5]
\begin{equation}
\Phi=\Psi=-\zeta+H{\cal I}\;,
\end{equation} 
(where $H(t)\equiv \dot{a}(t)/a(t)$), and in this mode there is no anisotropic inertia.  (Eq.~(5) can be written as the formula by which $\zeta$ is usually defined[6], $\zeta=-\Psi+\delta\rho/3(\bar{\rho}+\bar{p})$.)  Observation of the cosmic microwave background can tell us whether the cosmological perturbations are 
actually in this mode just before they re-enter the horizon, so it is of great interest to ask, in what cosmological models we would expect the perturbations to be purely adiabatic outside the horizon.

Before addressing this question, let's recall how current observations are used to check the 
adiabatic nature of the perturbations.  It appears that during the whole period when the cosmic temperature is in the range
$10^9\,{}^\circ{\rm K}\gg T\gg 3000\,{}^\circ{\rm K}$, the constituents of the universe consist of a non-relativistic nucleon-electron
plasma, plus photons, plus highly relativistic free neutrinos, plus non-relativistic cold dark matter.  To a good approximation, energy is separately conserved in each of these fluids, so each satisfies the energy conservation conditions
\begin{equation} 
\dot{\bar{\rho}}_\alpha+3H(\bar{\rho}_\alpha+\bar{p}_\alpha)=0
\end{equation} 
and, outside the horizon,
\begin{equation}
\delta\dot{\rho}_\alpha+3H(\delta\rho_\alpha+\delta p_\alpha)=3(\bar{\rho}_\alpha+\bar{p}_\alpha)\dot{\Psi}\;.
\end{equation} 
where $\alpha$ runs over the four fluids --- cold dark matter, neutrinos, photons,
and the electron-nucleon plasma.  For each of these fluids the pressure $p_\alpha$ is either negligible or a function only of the energy density $\rho_\alpha$, so it follows from Eqs.~(6) and (7) that[7]
\begin{equation}  
\frac{\delta\rho_\alpha}{\dot{\bar{\rho}}_\alpha}=
-\frac{\Psi+\zeta_\alpha}{H}\;,
\end{equation}  
where the $\zeta_\alpha$ are various constants.  (There are actually three separate constants
for the three neutrino flavors.) 
According to Eq.~(1), an adiabatic mode would have all $\zeta_\alpha$ equal.  In fact, 
according to Eqs.~(3) and (5), they would  all be equal  to $\zeta$.  The constants $\zeta_\alpha$ affect the further evolution of
the constituents of the universe after horizon re-entry and up to the time of last scattering, and so can be measured by 
observations of anisotropies in the cosmic microwave background.  So far, there is no evidence for unequal $\zeta_\alpha$, and some indications that they are equal for cold dark matter and photons.[8]

The most usually considered case in which we would expect purely adiabatic perturbations is inflation with a single scalar ``inflaton'' field.  It is well known that in this case the cosmic perturbations are purely adiabatic during inflation after horizon exit (whether or not the ``slow roll'' conditions are satisfied), and reference [1] shows that they then remain adiabatic through the period of reheating and as long as the perturbations remain outside the horizon.  In this case, all $\zeta_\alpha$
would be equal.
With several scalar fields the perturbations are not in general purely adiabatic during inflation, and it seems to be widely expected that this would show up in unequal values for the constants $\zeta_\alpha$.  

But there is a plausible scenario in which non-adiabatic fluctuations produced in multi-field inflation become adiabatic {\em after} reheating, and then remain adiabatic until horizon re-entry.  Suppose that reheating leaves the universe in a state of local thermal (including chemical) equilibrium, {\em with no non-zero conserved quantities.}  This is what is generally assumed in conventional theories of cosmological baryogenesis.  In this
case the perturbations are described by  just three degrees of freedom: the temperature fluctuation $\delta T$, the scalar potential $\Phi=\Psi$, and the velocity potential fluctuation $\delta u$.  As described in [3], these variables are governed by three first-order differential equations, so they  must have three independent solutions, but there is also a constraint relating $\delta T$, $\Psi$, and $\delta u$, leaving just two independent solutions.  Since there are always two independent adiabatic solutions, the solutions here are all adiabatic.   (This argument would not apply if one or more scalars remained decoupled from the matter in thermal equilibrium [9], as in curvaton models [10], theories of scalar field baryogenesis [11], and in some theories of cosmological moduli [12] or axions [13].)

Later the universe develops a non-zero baryon number; neutrinos and cold dark matter decouple from other particles; and local thermal equilibrium is lost.  The arguments of [3] then no longer apply, which is why the field equations in the era from $e^+e^-$ annihilation to last scattering have solutions (8) with different $\zeta_\alpha$.  Nevertheless, as long as the wavelength remains outside the horizon the adiabatic solution (2)--(5) is always a solution, and since the cosmological perturbations were described by this solution during the assumed early era of local thermal equilibrium, 
they remain purely adiabatic.  

Just as in [1], there is a weak point in this argument.  Local thermal equilibrium is never perfect, so there are always other degrees of freedom besides $\delta T$, $\Psi$, and $\delta u$, and hence other solutions of the field equations besides the adiabatic solutions.  We shall show in Section II that any non-adiabatic contributions to cosmological fluctuations will decay as the universe approaches a state of local thermal equilibrium with zero chemical potentials.  But even if these non-adiabatic solutions make only a tiny contribution to cosmological perturbations as this state is approached,  
how do we know that they do not grow rapidly again as local thermal equilibrium is subsequently lost, as for instance by the decoupling of dark matter particles, or as non-zero chemical potentials appear in cosmological baryosynthesis?

It seems that this question must be addressed on a case by case basis.  Here we will consider it in two familiar contexts: the survival of cold dark matter particles as they go out of local thermal equilibrium, treated in Section III , and the appearance of a non-zero chemical potential through the out-of-equilibrium decay of a massive exotic particle, considered in Section IV.  

\vspace{6pt}

\begin{center}
{\bf II. APPROACH TO ADIABATICITY}
\end{center}

\vspace{12pt}

Suppose that after inflation the dominant constituent of the universe is a heat bath in local thermal equilibrium with no chemical potentials, but that, as a remnant of multi-field inflation, there is also present a species of particles with number density $n$ that is not even approximately described by the adiabatic condition (2).  The rate of change of this number density in co-moving inertial frames will be some
function $Y(n,T)$ of $n$ and of the temperature $T$ of the heat bath\footnote{Here $u^\mu$ is the velocity four-vector, normalized so that $g_{\mu\nu}u^\mu u^\nu=-1$, with unperturbed components $\bar{u}^0=+1$ and $\bar{u}^i=0$.  Because outside the horizon spatial gradients can be neglected, it is unnecessary to state whether $u^\mu$ is the velocity four-vector of the particles with number density $n$, or of the heat bath energy-momentum tensor, or of the total energy-momentum tensor --- all we need to know about $\delta u^\mu$ is that $\delta u^0=\delta u_0=-\Phi$, which follows from the normalization condition on $u^\mu$.)}
\begin{equation} 
(n\,u^\mu)_{;\mu}=Y\;.
\end{equation} 
To zeroth order in perturbations this is
\begin{equation} 
\dot{\bar{n}}+3H\bar{n}=\bar{Y}
\end{equation} 
while to first order
\begin{equation} 
\delta \dot{n} +3H\delta n-3\bar{n}\dot{\Psi} =\delta Y+\Phi\bar{Y}
\end{equation}
For simplicity, we will assume that there were so many particle species in equilibrium that $n$ can be ignored in the evolution of  $\Psi$ and $T$.  According to the argument quoted in Section I, it follows that  $\delta T$ and $\Psi$ are given by the adiabatic conditions:
\begin{equation} 
\frac{\delta T}{\dot{\bar{T}}}=-{\cal I}\;,~~~~~\Phi=\Psi=H{\cal I}-\zeta\;.
\end{equation} 
It follows that the perturbation to the rate of change of $n$ takes the form
\begin{equation} 
\delta Y+\dot{\bar{Y}}{\cal I}=\frac{\partial \bar{Y}}{\partial \bar{n}}(\delta n+\dot{\bar{n}}{\cal I})\;,
\end{equation} 
the terms involving $\partial \bar{Y}/\partial \bar{T}$ canceling.
Adding the time derivative of ${\cal I}$ times Eq.~(10) to Eq.~(11) and using Eqs.~(12) and
(13) gives
\begin{equation} 
\frac{d}{dt}\Big(\delta n+{\cal I}\dot{\bar{n}}\Big)=\left(-3H+\frac{\partial \bar{Y}}{\partial \bar{n}}\right) \Big(\delta n+{\cal I}\dot{\bar{n}}\Big)\;.
\end{equation} 
This must be compared with the rate of decrease of $\bar{n}$ itself, given by Eq.~(10).  From Eqs.~(10) and (14) we have
\begin{equation} 
\frac{d}{dt}\ln\left|\frac{\delta n+\dot{\bar{n}}{\cal I}}{\bar{n}}\right|=\frac{\partial \bar{Y}}{\partial \bar{n}}-\frac{\bar{Y}}{\bar{n}}
\end{equation} 
The right-hand side may have any sign, but we know that when $\bar{n}$ is near the value $n_{\rm EQ}(T)$ that it would have in local thermal equilibrium at a temperature $T$, the rate of change of $\bar{n}$ due to all causes apart from the expansion of the universe --- decay, annihilation, etc. --- must take the form
\begin{equation} 
Y\rightarrow -\Lambda(T)\,\Big(n- n_{\rm EQ}(T)\Big)
\end{equation} 
where $\Lambda$ is some rate that must be positive in order that $\bar{n}$ in Eq.~(10) should be 
{\em attracted} toward $n_{\rm EQ}$ when the expansion of the universe can be neglected.  For $\Lambda(\bar{T})\gg H$ the expansion of the universe {\em can} be neglected, and so as the density $\bar{n}$ approaches the value it would have in thermal equilibrium, the right-hand side of Eq.~(15) approaches the negative value $-\Lambda(\bar{T})$, and the ratio $(\delta n+\dot{\bar{n}}{\cal I})/\bar{n}$, which can be taken as a dimensionless measure of the departure from adiabaticity, decays exponentially with rate $\Lambda(\bar{T})$.  We will see in Sections III and IV that this ratio is the quantity that sets the scale for possible inequalities in the observable quantities $\zeta_\alpha$.

We have seen that the non-adiabaticities associated with multi-field inflation become small if inflation is followed by a period of local thermal equilibrium with no non-zero chemical potentials.  
Now we have to see if they can be revived when the universe subsequently goes out of thermal equilibrium.

\vspace{6pt}

\begin{center}
{\bf III. DARK MATTER}
\end{center}

\vspace{12pt}

Next we consider the possible non-adiabatic contributions to cosmological perturbations that might be caused by 
the departure from local thermal equilibrium that occurs when the annihilation rate of heavy cold dark matter particles can no longer keep up with the rate that these particles would disappear in thermal equilibrium.[14]   

We assume here that all matter and radiation except the cold dark matter particles are in local thermal equilibrium with each other, and that this thermal bath dominates the energy density of the universe.  Then
the density $n$ of the cold-dark matter particles will again be governed by Eq.~(15).  As we have seen
in Section II, if there is a sufficiently long period when the rate $Y$ characterizing interactions of the cold dark matter particles with the thermal bath is much larger than the expansion rate, then
by some time $t_1$ the  
ratio $|\delta n+\dot{\bar{n}}{\cal I}|/\bar{n}$ will have become exponentially small.   
Eventually, as the temperature falls
below the dark matter particle mass, the fractional rate of decrease of $n_{\rm EQ}$ will become greater than $Y$, and  $n$ will consequently become greater than $n_{\rm EQ}$,
eventually leaving over a remnant of dark matter that has survived to the present.  The 
question is whether the small non-adiabatic perturbation at time $t_1$ is amplified while the dark
matter goes out of equilibrium.

Even when the cold dark matter density $n$ begins to differ appreciably from its equilibrium value $n_{\rm EQ}$, Eq.~(15) will still apply, but $Y$ will no longer be given by Eq.~(16).  
If we assume that dark matter particles are annihilated only in binary collisions, and created only in pairs, then $Y$ will be of the form
\begin{equation} 
Y(n,T)=-R(T)\Big(n^2-n_{\rm EQ}^2(T)\Big)\;,
\end{equation} 
where $R$ is a positive rate constant, equal to the average 
over the dark matter velocity distribution of the product of the dark matter particle velocity and the annihilation cross section.  Then Eq.~(15) becomes 
\begin{equation} 
\frac{d}{dt}\ln\left|\frac{\delta n+\dot{\bar{n}}{\cal I}}{\bar{n}}\right|=
-R(\bar{T})\,\left(\frac{\bar{n}^2+n_{\rm EQ}^2(\bar{T})}{ \bar{n}}\right)\;.
\end{equation} 
At late times $\bar{n}\gg n_{\rm EQ}(\bar{T})$, so the right-hand side approaches $-R(\bar{T})\bar{n}$.
The rate constant $ R(\bar{T})$ approaches a constant at low temperature, and $\bar{n}$ 
asymptotically goes as $a^{-3}$, so the time-integral of the right-hand side of Eq.~(18) converges
for large $t$.  Therefore at late times
\begin{equation} 
\frac{\delta n+\dot{\bar{n}}{\cal I}}{\bar{n}}\rightarrow
\left[\frac{\delta n+\dot{\bar{n}}{\cal I}}{\bar{n}}\right]_1\exp\left[-\int^\infty_{t_1}
R(\bar{T})\left(\frac{\bar{n}^2+n_{\rm EQ}^2(\bar{T})}{ \bar{n}}\right)\,dt\right]\;.
\end{equation}
We see that the departure from adiabaticity at late times is {\em less} than 
at the end of the period when the cold dark matter particles were in thermal equilibrium with everything else, though only by a finite factor. 

At late times $\dot{\bar{n}}\rightarrow -3H \bar{n}$ and $H{\cal I}=\zeta+\Psi$, so
Eq.~(19) may be written as in Eq.~(8):
\begin{equation} 
\frac{\delta \rho_D}{\dot{\bar{\rho}}_D}=\frac{\delta n}{\dot{\bar{n}}}\rightarrow -\left(\frac{\zeta_D+\Psi}{H}\right)\;,
\end{equation} 
with $\zeta_D$ now given by
\begin{equation} 
\zeta_D=\zeta+\frac{1}{3}\left[\frac{\delta n+\dot{\bar{n}}{\cal I}}{\bar{n}}\right]_1\exp\left[-\int^\infty_{t_1}
R(\bar{T})\left(\frac{\bar{n}^2+n_{\rm EQ}^2(\bar{T})}{ \bar{n}}\right) \,dt \right]\;.
\end{equation}
As we have seen, the quantity $\left[(\delta n+\dot{\bar{n}}{\cal I})/\bar{n}\right]_1$
is exponentially small if there is a sufficiently long period when the rate $Y$ characterizing interactions of the cold dark matter particles with the thermal bath is much larger than the expansion rate, and Eq.~(21) shows that then the departure of the observable quantity $\zeta_D$ from the value 
$\zeta$ that it would have for a purely adiabatic perturbation is even smaller.
\vspace{6pt}

\begin{center}
{\bf IV. BARYOGENESIS}
\end{center}

\vspace{12pt}

Here we will consider  the simple original model of cosmological baryogenesis[15],  with the later modification[16], that the out-of equilibrium decay of some heavy exotic particle $X$ produces a 
non-zero value of the quantity $B-L$, which is subsequently processed by  
$B-L$-conserving non-perturbative effects of the electroweak interactions[17] into some definite proportions of $B$
and $L$, depending only on the numbers of generations and scalar doublets[18].    It is assumed that at some time $t_1$ after inflation a nearly perfect local thermal equilibrium has been reached, with no chemical potentials, but that subsequently the actual rate of disappearance of the $X$ particle (and its antiparticle, if distinct) does not keep up with the rapid decrease that would occur in thermal equilibrium as the temperature falls below the $X$ mass.  In the original model [15] of direct baryosynthesis the $X$ particles were assumed to have distinct antiparticles, but with equal number densities $n_X$. In the case [16] of leptogenesis the $X$ particles are usually assumed to be identical with their antiparticles, and their density will again be denoted $n_X$.  
We will assume that there are so many particle species in thermal equilibrium that the  density of the $X$ particles makes a negligible contribution to the gravitational field and to the evolution of the temperature of the particles in equilibrium, so that the fluctuations in the temperature and metric are given by Eq.~(12).  Then the degree of non-adiabaticity of this number density will again be given by an equation like
Eq.~(15), in which $Y$ is now the rate of change of $n_X$ in co-moving inertial frames.  As before, this gives
\begin{equation}
\frac{\delta n_X+\dot{\bar{n}}_X{\cal I}}{\bar{n}_X}=\left(\frac{\delta n_X+\dot{\bar{n}}_X{\cal I}}{\bar{n}_X}\right)_1\exp\left[\int_{t_1}^t \left(\frac{\partial \bar{Y}}{\partial \bar{n}_X}-
\frac{\bar{Y}}{\bar{n}_X}\right)\,dt\right]\;,
\end{equation} 
a subscript 1 denoting the time $t_1$, when the decay of the $X$ particles has not yet begun to produce any appreciable net density of $B-L$.  But once the temperature falls below the $X$-particle mass and the disappearance rate of the $X$-particles falls below the expansion rate, the density  $n_X$ will be much larger than its equilibrium value, and Eq.~(16) will not apply.  

Suppose that the disappearance of a single $X$ particle (together with its antiparticle, if distinct)
produces on average a net value $b$ of $B-L$.  The rate of change of the density $n_{B-L}$ of $B-L$
in co-moving inertial frames is then
\begin{equation} 
(n_{B-L}u^\mu)_{;\mu}=-b\,(n_X\,u^\mu)_{;\mu}=-b\,Y(n_X,T)\;.
\end{equation} 
To zeroth order in perturbations, this gives 
\begin{equation} 
\dot{\bar{n}}_{B-L}+3H\bar{n}_{B-L}=-b\,\bar{Y}\;,
\end{equation} 
while to first order 
\begin{equation} 
\delta\dot{n}_{B-L}+3H\delta n_{B-L}-3\bar{n}_{B-L}\dot{\Psi}=-b\Big(\delta Y+\Phi\bar{Y}\Big)\;.
\end{equation}
The sum of  Eq.~(25) and the time derivative of ${\cal I}$ times Eq.~(24) is
\begin{equation} 
\frac{d}{dt}\Big(\delta n_{B-L}+\dot{\bar{n}}_{B-L}{\cal I}\Big)+3H\Big(\delta n_{B-L}+\dot{\bar{n}}_{B-L}{\cal I}\Big)=-b\,\frac{\partial \bar{Y}}{\partial \bar{n}_X}(\delta n_X+{\cal I}\dot{\bar{n}}_X)\;.
\end{equation}
The solution is 
\begin{equation} 
\delta n_{B-L}+\dot{\bar{n}}_{B-L}{\cal I}=b\,\left(\frac{a_1}{a}\right)^3\,\left[1-\exp\left(\int_{t_1}^t \frac{\partial \bar{Y}}{\partial \bar{n}_X}\,dt\right)\right]
\Big[\delta n_X+\dot{\bar{n}}_X{\cal I}\Big]_1\;,
\end{equation} 
a subscript 1 again denoting quantities evaluated at time $t_1$.  
Also, the solution of Eq.~(24) is
\begin{equation} 
\bar{n}_{B-L}=-b\left[\bar{n}_X-\left(\frac{a_1}{a}\right)^3\bar{n}_{X\,1}\right]\;.
\end{equation} 
Thus the departure from adiabaticity in the $B-L$ density is
\begin{equation} 
\frac{\delta n_{B-L}+{\cal I}\dot{\bar{n}}_{B-L}}{\bar{n}_{B-L}}=
\left(\frac{\left(\delta n_X+{\cal I}\,\dot{\bar{n}}_X\right)_1}{\bar{n}_{X\,1}-(a/a_1)^3\bar{n}_X}\right)
\left[1-\exp\int_{t_1}^t\frac{\partial\bar{Y}}{\partial \bar{n}_X}\,dt\right]
\end{equation} 
Asymptotically only decay contributes to the disappearance rate $Y$, so $\bar{Y}\rightarrow -\Gamma_X\bar{n}_X$  (where $\Gamma_X$ is the decay rate of the $X$-particle), and therefore the exponential in Eq.~(29) becomes negligible compared with 
one and $(a/a_1)^3\bar{n}_X$ becomes negligible compared with $\bar{n}_{X\,1}$.  Also, after
the temperature drops below the electroweak scale $\approx 300$ GeV, the
 values of the zeroth and first-order terms in the baryon density $\bar{n}_B$
and $\delta n_B$ are proportional respectively to $\bar{n}_{B-L}$ and $\delta n_{B-L}$ with the
same constant factor, so 
\begin{equation} 
\frac{\delta n_B+\dot{\bar{n}}_B{\cal I}}{\bar{n}_B}\rightarrow \frac{\delta n_{B-L}+{\cal I}\dot{\bar{n}}_{B-L}}{\bar{n}_{B-L}}\rightarrow \left(\frac{\delta n_X+{\cal I}\dot{\bar{n}}_X}{\bar{n}_X} \right)_1
\end{equation} 
Equivalently, since at late times $\bar{n}_{B}\propto a^{-3}$ and $H{\cal I}=\zeta+\Psi$,
\begin{equation}
\frac{\delta n_{B}}{\dot{\bar{n}}_{B}}\rightarrow -\left(\frac{\Psi+\zeta_{B-L}}{H}\right)\;,
\end{equation}
where $\zeta_{B}$ is the constant
\begin{equation} 
\zeta_{B}= \zeta +\frac{1}{3}\left[\frac{\delta n_X+\dot{\bar{n}}_X{\cal I}}{\bar{n}_X}\right]_1\;.
\end{equation} 
As we saw in Section II, if there is a sufficiently long period ending at time $t_1$ during which the $X$ particles are decaying and being recreated by the heat bath of other particles at a rate much greater than $H$, then  the second term on the right-hand side of Eq.~(32) will be exponentially small.  Hence, on these assumptions, the observed constant $\zeta_{B}$ will be very close to the common value $\zeta$ of the other $\zeta_\alpha$.

\vspace{6pt}

\begin{center}
{\bf V. CONCLUSION}
\end{center}

\vspace{12pt}

If we assume that the results of Sections III and IV are typical of the general case, then the existence of a post-inflation period of nearly perfect local thermal equilibrium would rule out any appreciable present departures from a purely adiabatic perturbation.  Thus the observation of a departure from  a purely adiabatic perturbation, such as a measurement of different values for the $\zeta_\alpha$ in Eq.~(8), would be evidence not only that non-adiabatic perturbations are generated in inflation, but also that there was no era
of perfect or nearly perfect local thermal equilibrium when all conserved quantum numbers vanished.

I am grateful for helpful conversations with E. Komatsu and M. Tegmark.
I also thank K. Chaicherdsakul for pointing out typographical errors in
an earlier version of this paper.  This material is based upon work supported by the 
National Science Foundation under Grant No. 0071512 and with support from the Robert A. Welch Foundation, Grant Nos. F-0014 and F-1099, and also grant support from  the US Navy, Office of Naval Research, Grant No. N00014-03-1-0639, Quantum Optics Initiative. 
\

\begin{center}
{\bf REFERENCES}
\end{center}

\nopagebreak

\begin{enumerate}

\item S. Weinberg, astro-ph/0401313.

\item See, e.g., C. Armend\'{a}riz-Pic\'{o}n, astro-ph/0312389.

\item S. Weinberg, Phys. Rev. D {\bf 67}, 123504 (2003); S. Bashinsky and U. Seljak, astro-ph/0310198.  Also see the appendix to S. Weinberg, astro-ph/0306374.  

\item  The existence of solutions with $\delta {\cal S}/\dot{\bar{\cal S}}$ equal for all energy densities, pressures, etc. (but not the detailed solution (2)--(5)) seems to have been generally accepted for a long time.   An intuitive ``separate universe''  argument for the existence of a solution satisfying Eq.~(1) has been given by D. H. Lyth and D. Wands, Phys. Rev. D {\bf 68}, 103516 (2003); also see D. Wands, K. A. Malik, D. H. Lyth, and A. R. Liddle, Phys. Rev. D {\bf 62}, 043627 (2000);  A. R. Liddle and D. H. Lyth, {\em Cosmological Inflation and Large-Scale Structure} (Cambridge University Press, 2000).   But this sort of argument only shows that there is a solution satisfying Eq.~(1) for zero wave number.  There are indeed many such solutions for zero wave number, most of which have no physical significance because they cannot be extended to finite wave number.  In ref. [3] it was shown that the requirement that the solution can be extended to finite wave number yields just two solutions, described by Eqs.~(2)--(5).  (It is this requirement
that requires that the infinitesimal re-definition of the time coordinate used in the ``separate universe'' argument to generate the solutions for zero wave number be accompanied with an infinitesimal re-scaling of the space coordinate.)
 
\item  D. Polarski and A. A. Starobinsky, Nucl. Phys. B {\bf 385}, 623 (1992), found the solution (1)--(5) for the field equations of two scalar fields interacting with each other only through gravity, but did not extend this result to general systems of particles and/or fields.  This explicit solution was not found by the ``separate universe'' arguments of Ref. 4.

\item J. M. Bardeen, P. J. Steinhardt, and M. S. Turner,
Phys. Rev. {\bf D28}, 679 (1983).  This quantity was 
re-introduced by D. 
Wands, 
K. A. Malik, D. H. Lyth, and A. R. Liddle, Phys. Rev. {\bf D62}, 
043527 
(2000).  Also see D. H. Lyth, Phys. Rev. D {\bf 31}, 1792 (1985); K. A. Malik, D. 
Wands, and C. Ungarelli, Phys. Rev. D {\bf 67}, 063516 (2003).

\item S. Bashinsky and U. Seljak, astro-ph/0310198.

\item H. V. Peiris {\em et al.}, Astrophys. J. Suppl. {\bf 148}, 213 (2003); P. Crotty, J. Garcia-Bellido, J. Lesgourgues, and A. Riazuelo, Phys. Rev. Lett.
{\bf 91}, 171301 (2003).

\item D. Polarski and A. A. Starobinsky, Phys. Rev. D {\bf 50}, 6123 (1994).

\item D. H. Lyth and D. Wands, Phys. Lett. {\bf B 524}, 5 (2002); D. H. Lyth, C. Ungarelli, and D. Wands, Phys. Rev. D {\bf 67}, 023503 (3003); D. H. Lyth and D. Wands, Phys. Rev. D {\bf 68}, 103516 (2003).

\item T. Moroi and M. Murayama, Phys. Lett. {\bf B 553}, 126 (2003).

\item T. Moroi and T. Takahashi, Phys. Lett. {\bf B 522}, 215 (2001).

\item D. Seckel and M. S. Turner, Phys. Rev. D {\bf 32}, 3178 (1985).

\item B.W. Lee and S. Weinberg, Phys. Rev. Lett. {\bf 39}, 165 (1977)

\item S. Weinberg, Phys. Rev. Lett. {\bf 42}, 850 (1979).  For earlier discussions of cosmological baryosynthesis, see A. D. Sakharov, JETP Lett. {\bf 6}, 24 (1967); M. Yoshimura, Phys. Rev. Lett. {\bf 41}, 281 (1978); {\bf 42}, 746(E) (1979);
S. Dimopoulos and L. Susskind, Phys. Rev. D {\bf 18}, 4500 (1979); Phys. Lett. {\bf 81B}, 416 (1979);
A. Yu. Ignatiev, N. V. Krosnikov, V. A. Kuzmin, and A. N. Tavkhelidze, Phys. Lett. {\bf 76B}, 436 (1978); B. Toussaint, S. B. Treiman, F. Wilczek, and A. Zee, Phys. Rev. D {\bf 19}, 1036 (1978); J. Ellis, M. K. Gaillard, and D. V. Nanopoulos, Phys. Lett. {\bf 80B}, 360 (1979); {\bf 82B}, 464(E) (1979).

\item M. Fukugita and T. Yanagida, Phys. Lett. B {\bf 174}, 45 (1986).

\item G. 't Hooft, Phys. Rev. Lett. {\bf 37}, 8 (1976); V. A. Kuzmin, V. A. Rubakov, and M. E. Shaposhnikov, {\em Phys. Lett.} {\bf 155B}, 36 (1985).

\item J. A. Harvey and M. S. Turner, Phys. Rev. D {\bf 42}, 3344 (1980).

\end{enumerate}
\end{document}